# Energy exchange in a bi-flux diffusion process consisting of particles of the same nature split into two distinct microstates

Luiz Bevilacqua[a], Maosheng Jiang[b]

[a]A.L.Coimbra Institute, COPPE-Federal University Rio de Janeiro, RJ, Brazil[1]
[b]School of Mathematics and Statistics, Qingdao University, Qingdao 266071, China

*In memory of Professor Leonardo Goldstein for his contribution to the advancement of mechanical science and engineering and strengthening of the Brazilian Society of Mechanical Sciences and Engineering*

Abstract

The present paper introduces a new approach to the dynamics of a particle system, split into two distinct microstates diffusing in a homogeneous medium. The particles belonging to the main microstate spread according to the classical Fick's law and the complementary set moves excited by a new potential. Each set is associated with a particular energy level. The particles can move between the two sets, introducing a third flux which is internal to the system. The governing equation is a fourth order PDE containing two new parameters, which can be time-dependent functions, in addition to the classical diffusion constant. It is shown that the solutions can avoid violations of the mass conservation requirements.

Keywords: Diffusion; Energy microstates; Fourth order PDE; Bi-flux theory; Entropy

**Introduction.**

Almost all diffusion processes are focused on the motion of a single class of particles or components that may move in a certain substratum excited by a single potential. Diffusion processes have shown to be adequate for modelling several types of phenomena ranging from physicochemical processes to socioeconomic behaviors. The most popular potential used to excite the motion of the elements in the process was proposed by Fick in the XIX century. Many experimental and theoretical works have contributed to confirm the adequacy of this type of process to simulate real events. Particularly important are the works focusing on the determination of the diffusion parameters in the governing equations.

Recently with the improvement of observation methods it has been possible to go deeper into physicochemical mechanisms and to explain more precisely certain peculiar behaviors. Particularly important are processes where particles may interact modifying internal properties that could interfere in the flux behavior. Diffusion processes involving two or more components spreading with different diffusion coefficients and responsive to intrinsic chemical reactions have been extensively studied. Reaction-diffusion equations are non-linear second order partial differential equations leading to remarkably interesting phenomena with very impressive pattern formations.

[1] bevilacqua@coc.ufrj.br ORCID 0000-0002-7695-1385

Despite the extraordinary development of new models to simulate physicochemical phenomena most of the new theories are second order linear or non-linear differential equations. The same is valid for other type of phenomena in socioeconomics and population dynamics, knowledge diffusion and disease transmission making use of diffusion models.

Fourth order diffusion equations have been used to model the behavior several physicochemical phenomena. Most of the models are non-linear, comprising reaction-diffusion terms. The evolution of the density distribution may present fluctuations leading to violations of thermodynamic principles and mass conservation requirements. A detailed analysis of such equations can be found in [1, 2]. A similar type of phenomenon can appear in the solution of the bi-flux equation introduced here. However, the phenomenological interpretation and the solution to solve the violations referred above follow a different orientation, as will be seen in the following sections. Particularly important is the effect of the time dependence of the new physical parameters on the behavior of the solution.

The purpose of the present paper is to introduce a new class of fourth order diffusion equation. The new governing equation shows that the particles are divided into two sets excited by two different potentials $\Psi_1$ and $\Psi_2$. Therefore, it is possible to say that the particles are split into two distinct energy states moving simultaneously according to two distinct diffusion processes. The diffusion parameters appearing in the derivation of the new equation require the secondary flow to be subsidiary to the primary flow. The primary flow is consistent with classical diffusion law.

The institution of the diffusion equation as proposed here does not require neither artificial inclusions of extra terms nor higher order expansion of the classical Fick's potential. It is only necessary to impose that a fraction of the particles is delayed as compared with the main set following the classical diffusion theory. Both terms in the governing equation appear naturally in the derivation and are equally important since they play specific complementary roles. The bi-flux process is consistent with the progressive formation of a sequence of distinct microstates required by the proposed evolution law. In the particular universe that will be defined in the next sections this means that the system initially excited in an ideal active energy state progressively collapses towards an ideal unrecoverable energy state.

Particularly important in this new theory is that it shows that it is always possible to avoid violation of physical requirements, provided that proper choices of the diffusion parameters are introduced.

**The new diffusion equation.**

Let us take the discrete approach to model the diffusion process with particles spreading in a homogeneous substratum. Consider the corresponding one-dimensional space divided into a sequence of elementary cells with length Δx as shown in the Fig.1. The process consists of transferring a fraction $\beta<1$ of the content of each cell in equal parts to the left and to the right neighboring cells. The complementary fraction $(1-\beta)$ is retained in the cell for the time interval Δt. At each time step Δt the process is repeated. The mathematical representation of this procedure is given by:

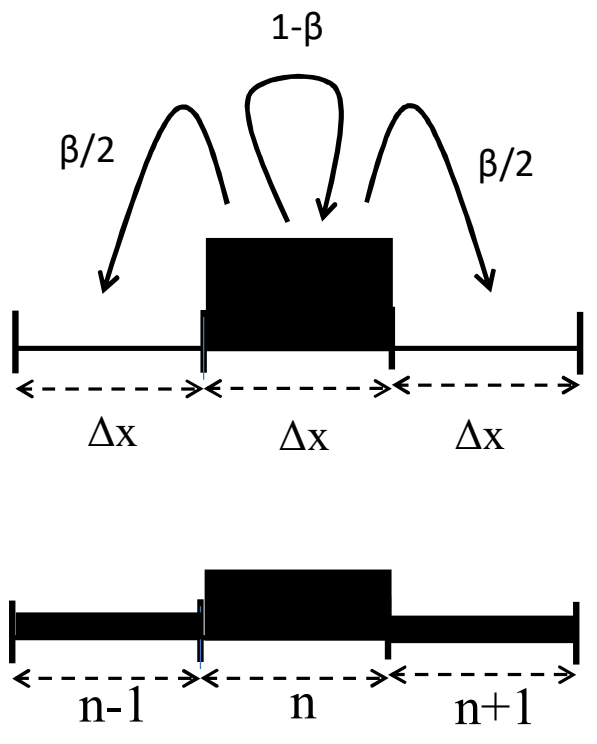

Fig.1. Distribution with partial retention. The fraction (1-β) remains trapped in the cell in the time interval Δt.

$$q_n^t = (1-\beta)q_n^{t-\Delta t} + \frac{1}{2}\beta q_{n-1}^{t-\Delta t} + \frac{1}{2}\beta q_{n+1}^{t-\Delta t}$$
$$q_n^{t+\Delta t} = (1-\beta)q_n^t + \frac{1}{2}\beta q_{n-1}^t + \frac{1}{2}\beta q_{n+1}^t$$

Considering the difference $\left(q_n^{t+\Delta t} - q_n^t\right)$ and after some algebraic operations the following difference equation is obtained:

$$\frac{\Delta q_n^{t+\Delta t}}{\Delta t} = \beta \left\{ D\frac{\Delta^2 q_n}{\Delta x^2} - (1-\beta) R \frac{\Delta^4 q_n}{\Delta x^4} \right\}^{t-\Delta t}$$

Where $\Delta x^2/2\Delta t = D$ and $\Delta x^4/4\Delta t = R$ are scale factors corresponding to the positions of the mean values of the associated density distributions $\beta q_n$ and $(1-\beta)q_n$ at a given time t. For the classical case it is well known that $\langle \bar{x} \rangle \propto t^{1/2}$ and for the new flux the mean value of the corresponding density distribution is related to time as $\langle \bar{x} \rangle \propto t^{1/4}$. The parameters $D$ and $R$ are therefore finite and independent.

Let $q(x,t)$ belong to the class $C^4$ with respect to the variable $x$ and $C^1$ with respect to the variable $t$, taking the limits $\Delta x \to 0$, $\Delta t \to 0$ and $\Delta q \to 0$ the new governing equation may be written as:

$$\frac{\partial q}{\partial t} = \beta D \frac{\partial^2 q}{\partial x^2} - (1-\beta)\beta R \frac{\partial^4 q}{\partial x^4} \tag{1}$$

The main characteristic of this new equation is to put in evidence the existence of two simultaneous fluxes. The particles scattering in the system are split into two sets. The first set corresponding to the fraction $\beta$ moves according to the classical Fick's law. The complementary set $(1-\beta)$ moves according to a new law derived from the second term on the right-hand side of equation (1). That is, equation (1) applies provided that there are two sets of particles sensitive to two distinct potentials, namely the classical Fick's potential $\Psi_1 = q(x,t)$ and a new potential given by $\Psi_2 = -\beta\partial^2 q/\partial x^2$. The respective fluxes, for a homogeneous media, are $\mathbf{\Psi_1} = -D\partial q/\partial x\,\mathbf{e_1}$ and by $\mathbf{\Psi}_2 = R\beta\partial^3 q/\partial x^3\,\mathbf{e_1}$.

It is remarkable that while the classical flux depends only on a material constant, namely the diffusion coefficient $D$, the secondary flux depends on a new parameter $R$ that we call reactivity factor and also on $\beta$ which represents the fraction of particles diffusing according to the classical law. This means that the secondary flux is subsidiary to the main flux, it exists if and only if the main flux is activated. Consequently, if there are no particles in the main flux, $\beta$ =0, then $\partial q/\partial t = 0$ the system becomes stationary [3, 4].

The new parameters are clearly associated with a new particle dynamic process intrinsic to the bi-flux phenomenon, namely, the internal flow of particles between two different energy states. Each flux is associated with a particular energy state and the particles tend to move from the upper energy to the lower energy state. Therefore, the equation proposed here introduces a new phenomenon, namely, the excitation of an internal flux. This third flux is expected to depend on the density $q(x,t)$ which means that the new parameters can also be density dependent functions, that is, in general we have $R=R(q(x,t)$ and $\beta=\beta(q(x,t)$. In this paper we will assume as a first approximation $R=R(t)$ and $\beta=\beta(t)$ as mean values in the space domain. The problem is therefore homogenized in the theory developed here. First the behavior of solutions with $R$ and $\beta$ constant will be briefly examined and then a more detailed exploration for the case with time dependent parameters.

**The hard bi-flux equation. Diffusion with time independent parameters.**

Consider the equation (1) defined in a homogeneous medium with time independent coefficients. Let us call the fourth order diffusion equation with $D$, $R$ and $\beta$ constant the hard bi-flux diffusion equation. The solution to this equation with homogeneous boundary conditions, may fail to represent real physical phenomenon introducing fluctuations in the density distribution leading to negative values of the density which is not possible for real diffusion processes [5]. Therefore, solutions of equation (1) with arbitrary choices of $D$, $R$ and $\beta$ may lead to violations of physic laws as observed by other authors. It is however possible to solve this problem with the proper interpretation of the new parameters.

The solution of the classical diffusion equation corresponding to an initial distribution q(x,0) highly concentrated at x=0 is always well behaved. This is not always true for the equation (1). The disturbance introduced by the secondary flux may lead to solutions which violate the mass conservation principle. The ill-behaved evolution of the concentration introducing fluctuations in the solution is to be attributed to the presence of the secondary flux in the diffusion process. Consequently, the new parameters, $R$ and $\beta$ play a fundamental role in the behavior of the bi-flux process. The critical parameter is the reactivity coefficient $R$ that controls the secondary flux intensity.

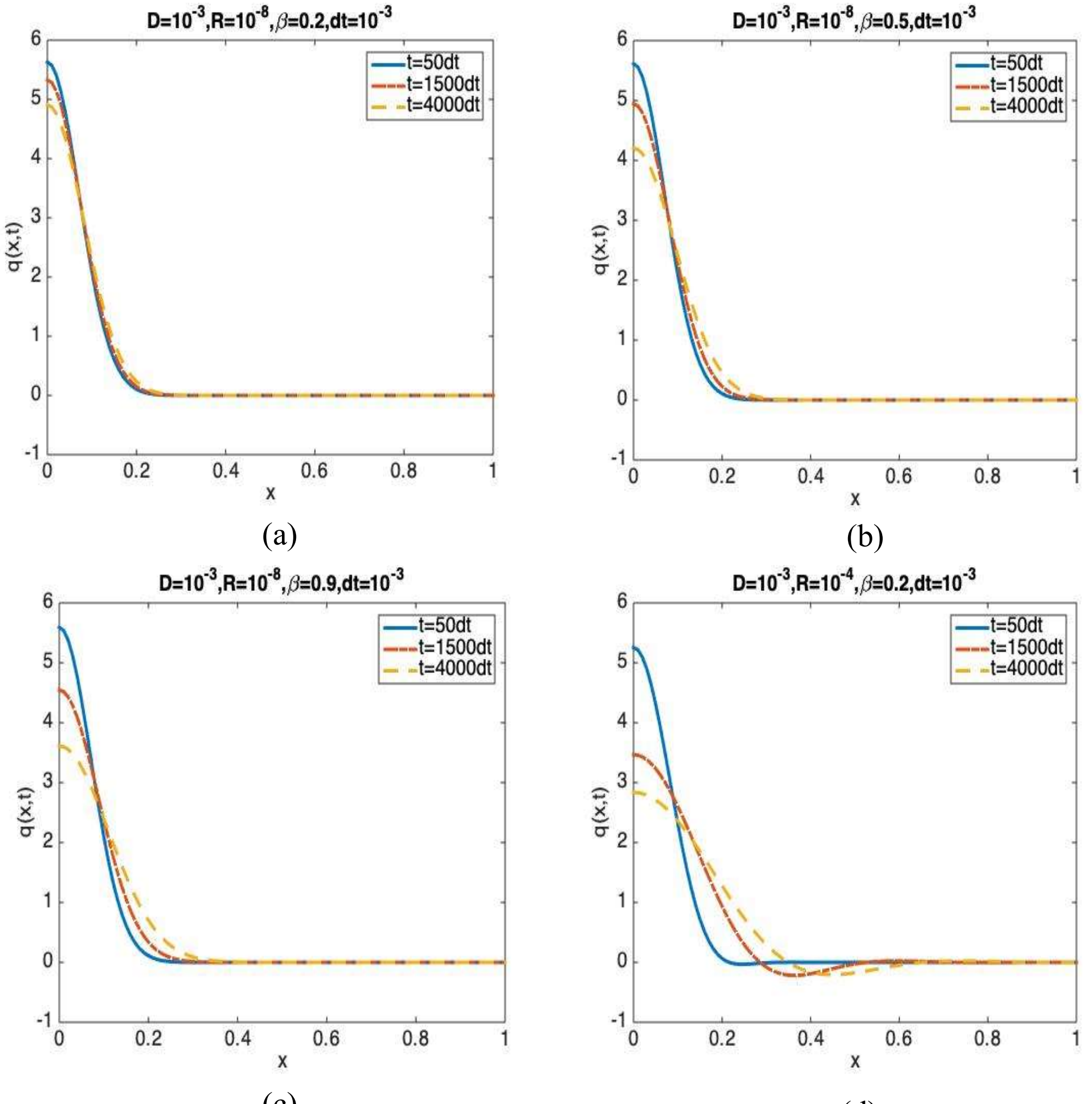

Fig 2. Evolution of the concentration profiles for different values of the reactiviy coefficient R and the fraction $\beta$ of particle in the primary flux: (a): $\beta$=0.2 R=$10^{-8}$;(b): $\beta$=0.5, R =$10^{-8}$; (c): $\beta$=0.9, R =$10^{-8}$; (d): $\beta$=0.2, R =$10^{-4}$ ; t is the time interval. For $R$ small the solution doesn't display no extremal(a), (b) and (c) irrespectively of the values of $\beta$. For large values of $R$ the solution presents extremal values and violates the mass conservation principle (d).

Consider the diffusion process defined in the domain [0,1] subjected to homogeneous boundary conditions. The initial condition is given by $q(x,0) = (\pi\varepsilon)^{-1/2} \exp\left(-x^2/\varepsilon\right), \varepsilon=0.01$. Let us take $D$ =$10^{-3}$ for all cases. Consider first the solutions for a constant reactive factor $R$=$10^{-8}$. The solutions are displayed in Fig.2 for $\beta$=0.2, $\beta$=0.5 and $\beta$=0.9. All the solutions, irrespectively of the values of $\beta$, do not violate the mass conservation principle. The concentration $q(x,t)$ remains positive for all $x$ and $t$. The distribution factor $\beta$ controls the speed but do not introduces fluctuations in the process. The higher the value of $\beta$ the faster the spreading of the solution in the medium.

Now if for $\beta$=0.2 we take the reactivity coefficient high enough, $R=10^{-4}$, the solution becomes incompatible with the mass conservation requirement. It would be only possible if an initial layer of particles were available to provide extra material necessary to supply the gaps created by the fluctuations in the process.

Solutions evolving at extremely high speeds display consistently at least one minimum as shown in Fig.2-d. For those cases, the concentration becomes negative for some subinterval $[x_1, x_2] \subset [0,1]$. This behavior has already been reported for several classes of fourth order parabolic equations. If the consideration of a secondary flux concomitant with the primary flux is missing it is difficult to find a plausible explanation.

The violation of the mass conservation law may be explained by the variations of the primary and secondary fluxes along the x-axis. For large values of $R$ the disturbances introduced in the primary and secondary fluxes are disruptive, both in intensity and direction. It is possible to induce reversal of primary flux recruiting particles from regions with low density and therefore not able to provide the number of particles required by the theory. It is not the purpose of the present paper to discuss this complex behavior that is left for future analysis.

**The soft bi-flux equation. Diffusion with time dependent parameters.**

The method used to derive the fundamental equation above does not incorporates possible time variations in the physical parameters. However particularly for adiabatic process it is expected $R$ and $\beta$ to be functions of time. Indeed, if the process occurs in an isolated environmental it is expected a progressive transfer of particles from the main flux to the secondary flux. Particles in the main Fickian process migrate progressively to the secondary energy state reducing the effective active energy and increasing the rotational kinetic energy. This process is admissible since the theory leads to the steady state as $\beta \to 0$ corresponding to the maximum rotational energy level. Therefore, consistently with this process both $\beta$ and $R$ must be functions of time. Note that $\beta$ and $R$ are interrelated as can be shown by solving the inverse problem [6, 7].

Let us take $\beta(t)=G(R(t)/R_0)$ where $R_0$ is a normalizing constant. The reactivity coefficient $R$ must be positive since the flux orientation is determined by the sign of $\partial^3 q/\partial x^3$ in the expression of $\boldsymbol{\Psi}_2$. Now since $0 \leq \beta \leq 1$ and $G(0)=1$ it is reasonable to assume $G(R/R_0)$ as a decreasing function of $R/R_0$ or equivalently, the fraction of particles belonging to the secondary state $(1-\beta)$ increases with $R$ up to a maximum value when the system reaches the steady state or more precisely freezes.

Now it is possible to highlight the most significant contribution characterizing the diffusion equation introduced here:

- *The dynamics of the particle motion given by equation (1) represents a twofold process namely the diffusion of two sets of particles excited by two distinct potentials and the simultaneous transfer of particles from the main energy state to the secondary degraded energy state.*
- *For irreversible processes – as assumed in this paper – the new parameters R and β are interrelated time dependent functions. The reactivity coefficient R controls the intensity of the secondary flux and the parameter β the transfer process between the two energy states.*

The equation (1) can now be written under a more appropriated form considering the diffusion process of two microstates interacting with each other in an isolated system:

$$\frac{\partial q}{\partial t} = \beta(t)D\frac{\partial^2 q}{\partial x^2} - (1-\beta(t))\beta(t)R_0 G^{-1}(\beta(t))\frac{\partial^4 q}{\partial x^4} \tag{2}$$

The fraction of particles in any of the two states does not affect the main flux intensity, but the fraction of particles $\beta$ in the main state is crucial to determine the secondary flux intensity. It is therefore easy to see that according to the present theory, an irreversible diffusion process tends to the freezing point, $\beta \to 0$, corresponding to the maximum rotational energy state where the active energy vanishes. All particles occupy a fixed position in the space domain.

We will explore the energy evolution in an ideal universe in the next section to explain the theory presented in this section.

**Energy exchanges in a particular universe.**

Let us assume an ideal universe consisting of a large number of particles, divided into two distinct energy microstates $E_1$ and $E_2$ scattering in an isolated system, interacting with each other and therefore exchanging energy. In our universe the total energy consists of kinetic energy divided into the two fundamentals forms, namely translational energy, and rotational energy. Any particle may be moving with linear momentum **p** and angular momentum **L**, the total energy $e = \mathbf{p}^2/2m + \mathbf{L}^2/2m\bar{d}$ however remains constant that is $e = e_p + e_\omega = constant$.. The active or observable energy consists of the translational energy. The rotational energy $P_\omega \propto \left(\mathbf{L}^2/2m\bar{d}\right)$ cannot be detected in our universe.

The hypothesis above is consistent with the equation (2) with the particles split into two sets $\beta$ and $(1-\beta)$, requiring the secondary flux $\mathbf{\Psi}_2$ to be subsidiary to the primary flux $\mathbf{\Psi}_1$. Indeed, assume that all particles excited in the state $E_1$ do not rotate, all kinetic energy is stored as active energy, while the kinetic energy corresponding to the state $E_2$ is stored as active energy associated to the flux $\mathbf{\Psi}_2$ and the rotational energy $P_\omega(\beta)$ as well.

The diffusion process introduced here assumes that at the very beginning all particles belong to the energy state $E_1$ where all particles are excited exclusively with translational motion associated to the flux $\mathbf{\Psi}_1$, that is $E_1 = f_1(|\mathbf{\Psi}_1|)$. As the process progresses the particles interact with each other exchanging energy and exciting other energy states consisting of translational and rotational energies. The translational observable energy corresponding to the second energy state is given by the secondary flux $\mathbf{\Psi}_2$. The hidden energy is the rotational energy $P_\omega(\beta)$. Therefore, the total energy corresponding to the secondary flux is given by $E_2=f_2(|\mathbf{\Psi}_2|,P_\omega)$.

Suppose that we have a system consisting of $N$ particles divided into two subsets, $N_1=\beta N$ and $N_2=(1-\beta)N$. The energy density or specific energy (energy/volume) corresponding to $N_1$ and $N_2$ can be defined respectively as:

$$U_1=\beta|\mathbf{\Psi}_1|^2/2q$$

and

$$U_2=(1-\beta)\left(|\mathbf{\Psi}_2|^2/2q+P_w(\beta)\right)$$

Now let $U_1$ and $U_2$ be the specific energies corresponding to the energy states $E_1$ and $E_2$ respectively. Let us introduce the following rules that apply to our particular universe.

1. *In an isolated system the total specific kinetic energy is given by* $U_1+U_2=\beta|\mathbf{\Psi}_1|^2/2q+(1-\beta)\left(|\mathbf{\Psi}_2|^2/2q+P_\omega(\beta)\right)$
   *Which is a time independent function.*
2. *In an isolated system, the active energy component in $U_1$ and $U_2$ corresponding to the flux potentials $\Psi_1$ and $\Psi_2$ decrease steadily as time increases, consequently the fraction $\beta$ necessarily decreases such that* $\lim_{t\to\infty}\beta(t)\to 0$
3. *Since the system preserves the total kinetic energy, translational and rotational, from the first two rules it is possible to write, $U_1+U_2=P_\omega(0)$*

Consider now the diffusion process active in the interval $x \in [a, b]$. The system is isolated and according to the second rule the total active energy tends to zero as $t \to\infty$ consistent with $\lim_{t\to\infty}\beta(t)\to 0$. Therefore, the linear momentum tends uniformly to zero for all $x$ as $t\to\infty$, that is $\lim_{t\to\infty}\mathbf{p}(x,t)\to 0$. Consistently with the conditions above let us find an expression for the variation of the fraction $\beta$ with the expected $P_\omega(\beta)$, the rotational energy production.

Clearly, the probability of interaction among particles inducing a reduction of particles in state $E_1$ is proportional to $\beta$. That is, the variation $\delta\beta$ is proportional to $\beta$. This means that the decay of translational or active energy into rotational energy ($\mathbf{p}\to\mathbf{L}$), is more intense when the number of particles in state $E_1$ is large, $\beta>>0$. The variable $P_\omega$ vanishes for $\beta=1$ and grows steadily as $\beta\to 0$ similarly to the reactive factor $R$ controlling the secondary flux intensity. Therefore, it is admissible to take $R$ also as measure of the rotational energy.

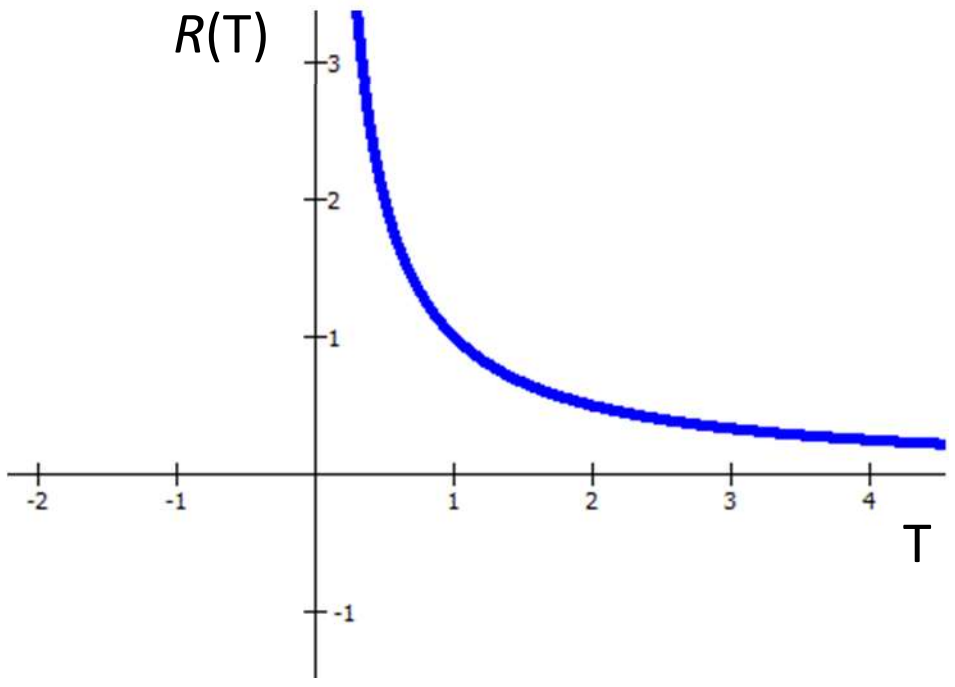

Fig3. Variation of the rotational energy with the parameter T

Let $T$ be a variable that measures the intensity of the active energy, such that as $T$ increases the active energy also increases. Given that the active energy is a function of time then $T = T(t)$ is also a function of time. Since the system is isolated and the total energy is constant, $T(t)$ is also an indirect measure of the rotational energy $R \approx P_\omega$ which is complementary to the active energy and therefore a decreasing function of $T$ (Fig.3).

At large active energy levels $T >> 0$ the rate of variation of the rotational energy $R(T)$ with respect to the variable $T$ is low. However, for $T$ small, big variations of the rotational energy occurs for relatively low decrease of the energy parameter $T$. Therefore, it is reasonable to assume that:

$$R(T) = \left(\frac{T_0}{T}\right)^n R_0 \tag{3}$$

Where $R_0$ and $T_0$ are normalizing constants. Let us take n=1. The variation $\delta\beta$ depends therefore on two determinant parameters:

(*1*) *The fraction of particles in the main energy state given by* $\beta$.
(*2*) *The variation of the rotational energy given by* $-\delta(R(T))$. *The negative sign meaning that* $\delta\beta$ *decreases as* $\delta R$ *increases.*

With the hypotheses above it is possible to define the variation of $\beta$:

$$\delta\beta = \beta\frac{(-\delta R)}{R_0}$$

From which follows:

$$\beta = \exp\left(-\frac{R}{R_0}\right) \tag{4}$$

Therefore $G(R/R_0) = \exp(-R/R_0)$ or $R = -R_0 \ln\beta$. The reactivity coefficient $R$ is a measure of the distribution of the particles between the two fundamental energy states $E_1$ and $E_2$. For $R$ equal zero and consequently vanishing rotational energy the active energy reaches the maximum possible level. This state corresponds to the maximum fraction of the active energy that can be converted into work. For $R\to\infty$ the system approaches a "dead state", maximum degradation, with the active energy being progressively transferred to an unrecoverable hidden energy state. Therefore, the hidden energy state consists of particles whose active energy is progressively and permanently reduced in the process. The relative distances among all particles tend to remain fixed and the system approaches a stationary state. All the energy is stored as rotational energy, the system is at rest meaning that $x_i$ = constant, i = 1,2...N, in a given reference frame. If it would only be possible to measure the active energy, that is, translation, then for very large values of $R$ an external observer would assume that the system is inactive or "dead". Maybe only the mass could be detected, and the rotational energy stored in the system would be hidden, it would be a kind of "non-observable energy". The reactivity coefficient $R$, is a measure of the rotational energy amount introduced in this paper, it is the counterpart of the entropy as defined in classical thermodynamics. Let:

$$\frac{\delta R}{R_0} = -n\left(\frac{T_0}{T}\right)^{n+1} \frac{\delta T}{T_0}$$

For n = 0 we have $\delta R$=0 and the rotational energy doesn't come into play. This hypothesis corresponds to the classical Fick's theory with active energy preserved. For n=1 the rotational energy is different from zero and refers to the energy state $E_2$ steadily growing over time. For $T_0/T >> 1$ small variations of $T$ imply very large variation of $\delta R$. That is for $T$ close to zero the rotational energy varies extremely fast with small variations $\delta T$.

**The fundamental equation for the simultaneous diffusion of two sets of particles interchanging energy in an irreversible process.**

According to the previous section, the fourth order equation (2) may be written in a more consistent manner regarding the energy distribution in the different microstates. Indeed, the partitions $\beta$ and $(1-\beta)$ are the probabilities $p_1$ and $p_2$ that particles belong to the energy state $E_1$ or $E_2$ respectively. The reactive coefficient $R$ is intrinsically associated with $\beta = p_1$ as proposed in the precedent section. Rewriting equation (2) with the help of the new parameters, we have:

$$\frac{\partial q}{\partial t} = p_1 D \frac{\partial^2 q}{\partial x^2} + p_2 R_0 p_1 \ln p_1 \frac{\partial^4 q}{\partial x^4} \quad (5)$$

With $S_1 = R_0\left(-p_1 \ln\left(p_1\right)\right)$ the equation (5) reads:

$$\frac{\partial q}{\partial t} = p_1 D \frac{\partial^2 q}{\partial x^2} - p_2 S_1 \frac{\partial^4 q}{\partial x^4} \quad (6)$$

Note that $S_1/R_0 = -p_1 \ln p_1$ is the definition of the Shannon entropy.

Recall that the flux intensity corresponding to the degraded microstate for a given variation of the density distribution $\partial q^3 / \partial x^3$ is determined by $\beta R = p_1 R = -R_0 p_1 \ln p_1$. Therefore, the variable $S_1$ may be associated to the rate of degradation characterizing the evolution towards the extinction of the active energy. The physical role of $S_1$ in the diffusion equation helps the understanding of the rotational energy evolution according to the statistic mechanics definition and information theory principles. The equation (6) says that the transfer of particles from the main energy state to the secondary energy developing in an isolated system follows an optimum path.

Closing this section let us assume that $R$ increases exponentially with time and $R(0) = 0$, that is, the system is initially at the maximum possible active energy state:

$$R(t) = R_0\left(\exp(\alpha t) - 1\right) \qquad (7)$$

Where $\alpha$ is a constant characteristic of the dissipation process. The distribution parameter or the probability of the system to be in the maximum convertible energy state, $\beta = p_1$ is then given by:

$$p_1 = \exp\left(1 - \exp(\alpha t)\right) \qquad (8)$$

and

$$S_1 = -R_0\left(1 - \exp(\alpha t)\right)\left[\exp\left(1 - \exp(\alpha t)\right)\right] \qquad (9)$$

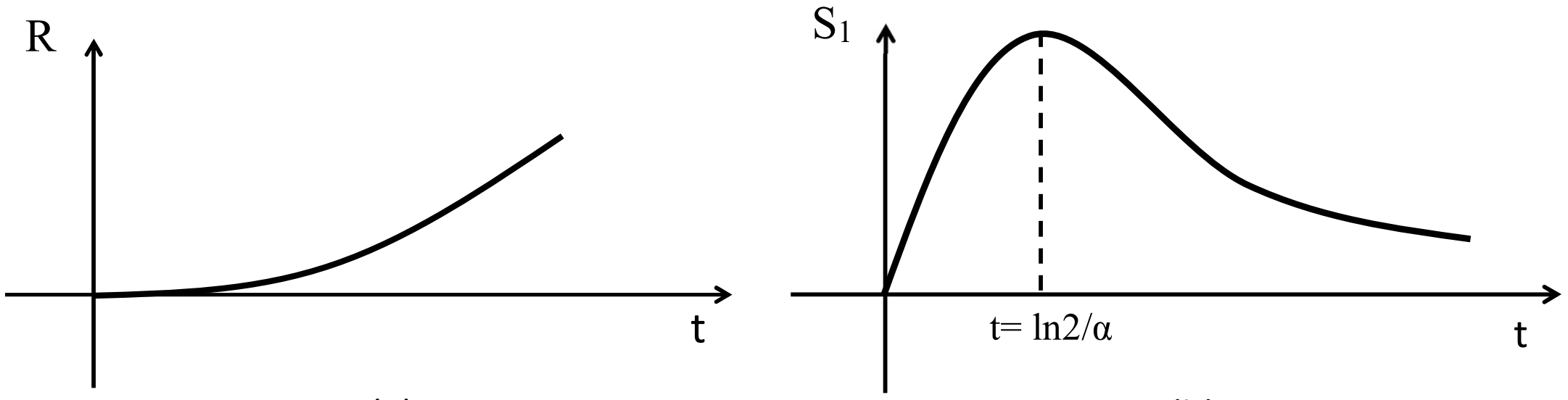

(a) (b)

Fig.4. ( a) Time evolution of the reactivity coefficient $R$ and (b) rate of degradation $S_1$ for a given constant α.

The process reaches its maximum degradation rate, max $S_1$, for t = ln2/α, Fig.4. The smaller the constant α the later the process reaches its maximum activity. The time variation assumed for $R$ is compatible with the expected evolution of the rotational energy in the system. The reactivity coefficient $R$ is an increasing function of time and the exponential function is flexible enough to match most of the regular physicochemical reactions. The variable $S_1$ derives directly from the definition of $R$ and is also compatible with the expected evolution behavior of the corresponding reaction. The process starts with a negligible activity and evolves gradually till a maximum when the degradation speed reaches a maximum. Thereafter the speed decreases gradually approaching eventually the frozen state.

**Examples of diffusion processes governed by the soft diffusion equation.**

Let us show that the introduction of the correct formulation to solve the bi-flux diffusion where the total internal energy is conserved is consistent with the mass preservation requirements. Consider the diffusion problem in the interval $x \in [-1,1]$ and $t > 0$. The system is isolated from the surroundings, therefore the boundary conditions correspond to no flux, primary and secondary, at both ends, $\mathbf{\Psi}_1\big|_{\pm 1} = 0$ and $\mathbf{\Psi}_2\big|_{\pm 1} = 0$. The initial condition is given by:

$$q(x.0) = 5\exp\left(-100x^2\right) \quad (10)$$

Now consider two cases corresponding to distinct physical properties. For the first one the parameters $\beta$ and $R$ are constants: $D = 0.2$, $R = 0.02$ and $\beta = 0.5$. For the second case the diffusion coefficient is constant as in the first case $D = 0.2$ but the fraction $\beta = p_1$ and the reactivity coefficient $R$ are time dependent functions as given in equations (8) and (9). The solutions are shown in the Fig. 5 for $R_0 = 0.02$ and α=1.

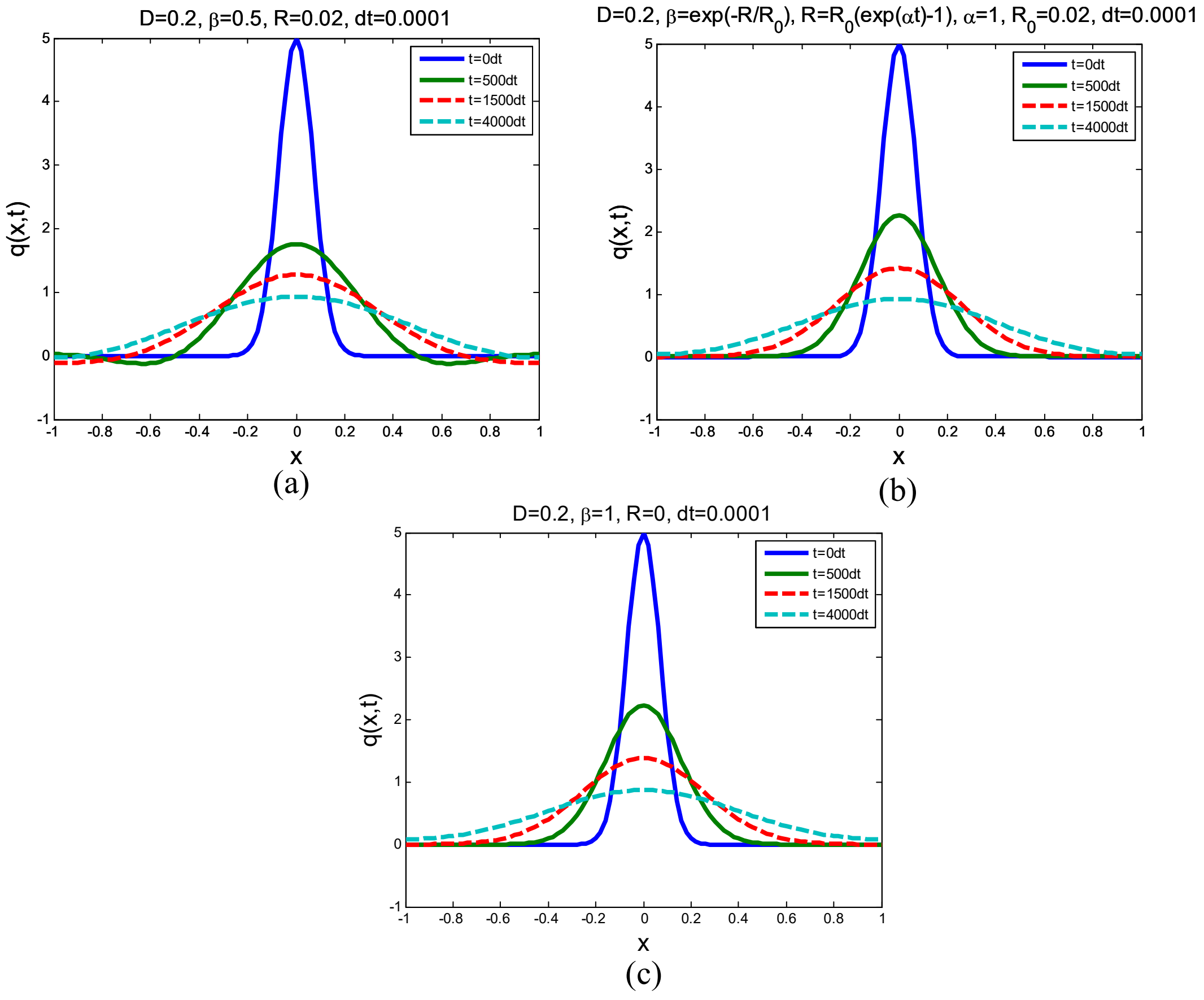

Fig.5. Time variation of the concentration distribution on [−1,1] for three cases;
(a) R and $p_1$ constants, (b) R function of time $R=R_0(\exp(\alpha t)-1)$ and $p_1 = \exp(-R/R_0)$,
(c) Classical Fick's diffusion process

The solution for the first case exhibits negative values of the concentration as shown in Fig.5-a. Therefore, this case would only be meaningful with the presence of an external source providing a sufficient number of particles to avoid flux disruption. This is equivalent to say that the system could not be isolated, for the prescribed initial conditions, violating the basic assumptions.

For isolated systems as considered above the solution to be consistent requires $R$ and $p_1$ to be functions of time since the rotational energy is not stationary. The solution for $R$ given by equation (7) is displayed in the Fig. 5-b. Negative values are avoided in the interval $x \in [-1,1]$ and therefore the equation (1) applies. Fig.6-c shows the time variation of the concentration $q(0,t)$ for three cases. For the first case with $R$ and $\beta$ constant the process develops faster as compared with the classical Fick's diffusion, $R$=0. For the case where $R$ and $\beta$ are functions of time the diffusion rate slightly decreases. The solutions show that irreversible physicochemical processes can be consistently represented by the bi-flux equation provided that the reactivity coefficient increases gradually with time. This means, according to our proposal, that particles in the primary flux, microstate $E_1$, are steadily migrating to the secondary flux, microstate $E_2$, and consequently increasing the rotational energy active in the system.

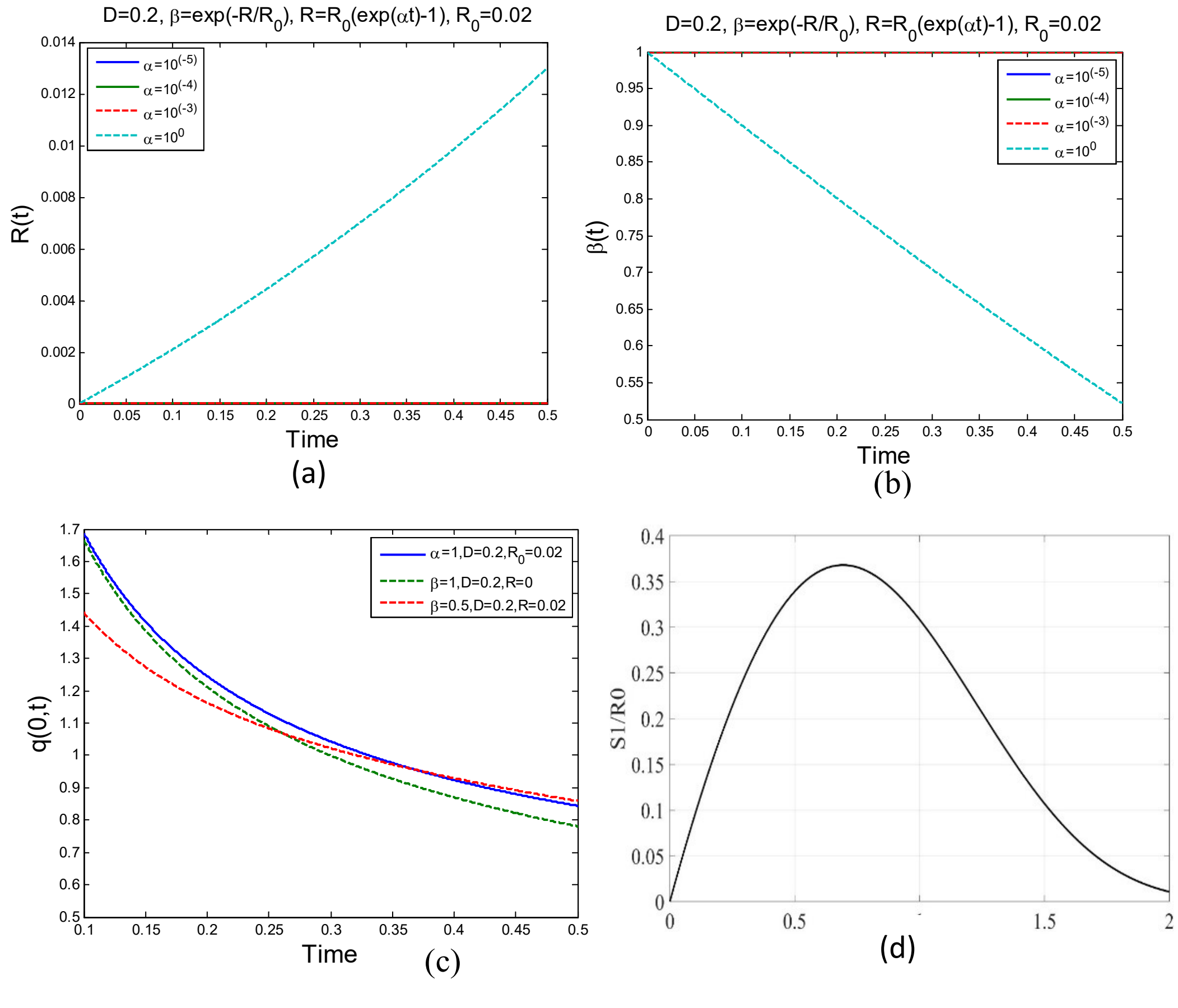

Fig.6. (a) and (b)Time variation of $R$ and $\beta$ for the case of time dependent parameters. (c) Time variation of the concentration at x=0 for the representative cases: classical Fick's diffusion, diffusion process with $R$ and $\beta$ constants and diffusion processes with $\beta$ and $R$ functions of time as given by equations (8) and (9). (d) Time variation of $S_1$ for the case of time dependent parameters associated to the speed of the degradation process.

The reactivity factor $R$ as given by equation (7) and the corresponding fraction $p_1$ are also displayed. The solution for the bi-flux equation is particularly sensitive to the ratio R/D for initial conditions consisting of particles highly concentrated at x=0. Negative values of the concentration are associated with high values of R/D. With R=R(t) taken as function of time, starting with R(0)=0, it is possible to control the time variation for R such that the solution behaves within the required positivity conditions. We may say that nature optimizes this process.

## Conclusion

This paper introduces a new diffusion equation representing the concomitant flux of two sets of particles of the same nature but moving under the action of two distinct potentials. The process develops in a particular universe where the kinetic energy prevails stored into two main states, namely translational and rotational states. The translational energy can produce work and was called active energy, but the rotational energy corresponds to an idle state. Particles storing exclusively rotational energy are fixed in a static configuration unable to produce work. With the theory introduced here, for adiabatic processes, where the total energy is preserved, it is possible to consider the transfer of active energy to rotational energy, leading eventually to a stationary state with the rotational energy reaching a maximum possible value. Even though the theory concerns an ideal universe it is adequate and frequently necessary to model several real cases where change of state is an important factor in the process. Particles may switch fluxes particularly in models simulating population dynamics and capital flow. The classical theory cannot describe properly these complex phenomena.

The problem posed by the violation of the mass conservation principle can be solved by the proper definition of the new parameters introduced in the new formulation. The fundamental hypothesis supporting the discrete approach used to derive the new equation leads clearly to the split of the flux into two streams dispersing at different rates. The main flux follows the classical Fick's law whereas the second flux is clearly dependent on the particle density in the primary flux. It is remarkable that this important correlation helps understanding the energy exchange taking place in the process.

The outbreak of fluctuations in the solution is due to the interference of the secondary flux in the process together with perturbations in the primary flux. For highly concentrated initial distributions close to x=0 the primary and the secondary fluxes, both contributes to the dispersion process leading to an accelerated reduction of particles close to the origin. But the secondary flux inverts sign at a relatively short distance from x=0 recruiting particles towards the origin. But since the initial removal of particles close to the origin is very intense at the beginning there are not enough particles to comply with the demand imposed by the secondary flux leading to negative values of the concentration.

To avoid disruptions in the density distribution it is necessary to interfere in the secondary flux. Since the secondary flux intensity depends on the product $\beta R$ the proper choice of these parameters may avoid formation of fluctuations and growth of negative values of the density function as well. With the proper choices of $\beta(t)$ and $R(t)$ the secondary flux develops slowly, and the density distribution allows for recruiting particles, inverting the flux direction, without disruption of the particle distribution. It is observed that for compatible bi-flux processes the value of $q(0,t)$ decreases slowly than for the classical Fick's diffusion process.

It was shown that the secondary flux intensity for the bi-flux soft equation, with $R$ and $\beta=p_1$ both suitable functions of time, depends on $S_1$ ($S_1 = -R_0 p_1 \ln p_1$). This parameter controls the secondary flux intensity and indirectly the rotational energy growth in the system. Since $S_1$ is the Shannon entropy it is possible to say that the diffusion process with change of energy states follows an optimum path. Nature optimizes the rotational energy growth in adiabatic processes. This is a remarkable outcome disclosed by the present development.

The approach proposed in this paper may be particularly useful in dissipation processes of living particles that could change states due to the liberation of internal energy [8]. The interaction with the surroundings with the interference of sources or sinks open a large spectrum for applications. Besides physicochemical phenomena the new equation may be helpful for the analysis of complex problems in population dynamics [9-12], socioeconomics dynamics [13-15], capital flow, knowledge flow [16] spreading of contagious diseases [17], and other fields where the hypothesis of a single flux is not satisfactory. Interaction with the surroundings, spreading in nonhomogeneous media and inclusion of reactive terms are necessary to deal with a large range of problems requiring coupling effects with other agents.

The bi-flux equation opens a very appropriate channel to deal with interactions between the particles in the diffusion process and external factors. Several cases of particles sensitive to particular disturbances in the dispersing medium can be solved with the bi-flux equation. The presence of a pheromone in a restrict region of the dispersing medium is a typical example [18]. The fourth order term is introduced to take into account the disturbance introduced by the external player in a restrict region. This type of problem can be modeled with the generalized equation:

$$\frac{\partial q}{\partial t} = \beta \frac{\partial}{\partial x}\left( D\frac{\partial q}{\partial x} \right) - (1-\beta)\frac{\partial}{\partial x}\left( \beta R \frac{\partial^3 q}{\partial x^3} \right)$$

Where $R=R(x)$ represents the presence of the external disturbance in the corresponding domain $[x_1, x_2]$.

Problems related to capital flow are also best modeled by using the present approach. In fact, the bi-flux process allows capital to flow in and out simultaneously in a given industrial complex as the result of selling products and buying supplies. The process represented by the bi-flux equation is controlled particularly by the $R/D$ ratio and the capital fraction $\beta$ at the entrance. If $\beta$ is small, which means that the volume of capital (not the flux intensity) in the inflow is less than the complementary fraction in the outflow, even so it is possible to have a positive result with the increase of capital accumulation. The necessary condition is to reduce $R$, which means improving the technology by reducing the internal transfer of capital from the inflow to the outflow needed to import expensive products. Although this is a well-recognized correlation, it is a primary test of the equation applied to quite simple problems. It simply says do not buy knowledge; internally generate the knowledge you need.

The bibliography was restricted to a relatively small number given that the present approach is very recent, and it was avoided introducing artificial references. In summary we are proposing a new tool that certainly needs improvements and adaptations regarding each application but open a wide window for more complete analysis of complex phenomena.

Ackowledgments

We are in debt with the IEA/USP for the support given to first author as a visiting researcher 2017/18, with the FINEP agency for the grant awarded to COPPE-UFRJ through the COPPETEC contract 21500 and with the CNPq for the continued support through the senior research grant.

# Appendix

**Symmetric distribution with retention**

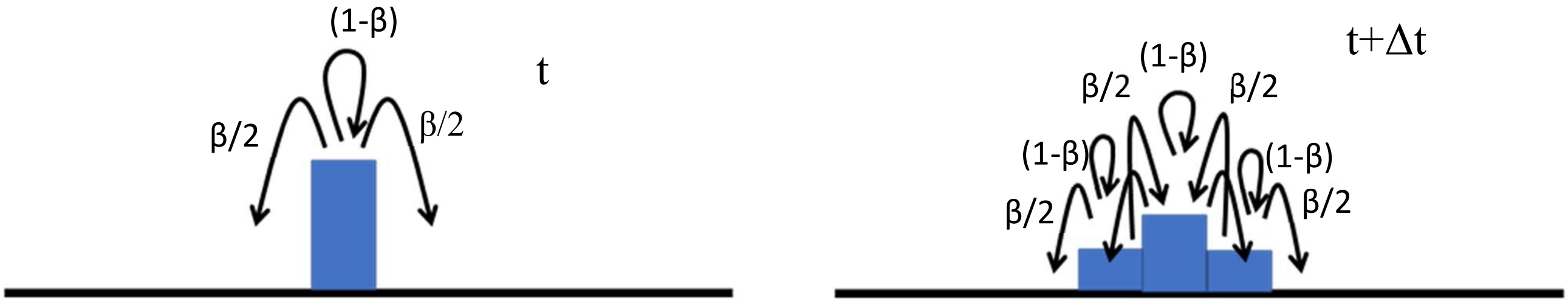

Fig. A-1. Contents distributions process with partial retention and symmetric redistribution

Consider the symmetric distribution with retention as displayed in Fig.A-1. The distribution scheme can be written as:

$$p_n^t = (1-\beta)p_n^{t-1} + \frac{1}{2}\beta p_{n-1}^{t-1} + \frac{1}{2}\beta p_{n+1}^{t-1} \qquad \text{(A1-a)}$$

$$p_n^{t+1} = (1-\beta)p_n^t + \frac{1}{2}\beta p_{n-1}^t + \frac{1}{2}\beta p_{n+1}^t \qquad \text{(A1-b)}$$

Reduction of the right-hand side of Eq. (A1-b) to time t-1 leads successively to:

$$p_n^{t+1} = (1-\beta)\left[(1-\beta)p_n^{t-1} + \frac{1}{2}\beta p_{n-1}^{t-1} + \frac{1}{2}\beta p_{n+1}^{t-1}\right] + \frac{1}{2}\beta\left[(1-\beta)p_{n-1}^{t-1} + \frac{1}{2}\beta p_{n-2}^{t-1} + \frac{1}{2}\beta p_n^{t-1}\right] +$$
$$\frac{1}{2}\beta\left[(1-\beta)p_{n+1}^{t-1} + \frac{1}{2}\beta p_n^{t-1} + \frac{1}{2}\beta p_{n+2}^{t-1}\right]$$

$$p_n^{t+1} = (1-\beta)^2 p_n^{t-1} + \beta(1-\beta)\left(p_{n-1}^{t-1} + p_{n+1}^{t-1}\right) + \frac{1}{4}\beta^2\left(p_{n-2}^{t-1} + p_n^{t-1} + p_{n-1}^{t-1} + p_{n+2}^{t-1}\right)$$

Subtracting $p_n^t$ given by (A1-a) it is obtained successively:

$$p_n^{t+1} - p_n^t = -\beta(1-\beta)p_n^{t-1} + \beta(1-\beta)\left(p_{n-1}^{t-1} + p_{n+1}^{t-1}\right) - \frac{1}{2}\beta\left(p_{n-1}^{t-1} + p_{n+1}^{t-1}\right) +$$

$$+\frac{1}{4}\beta^2\left(p_{n-2}^{t-1} + 2p_n^{t-1} + p_{n+2}^{t-1}\right)$$

$$p_n^{t+1} - p_n^t = \beta\left[-\frac{1}{2}\left(p_{n-1}^{t-1} + p_{n+1}^{t-1}\right) + (1-\beta)\left(p_{n-1}^{t-1} + p_{n+1}^{t-1} - p_n^{t-1}\right) + \frac{1}{4}\beta\left(p_{n-2}^{t-1} + 2p_n^{t-1} + p_{n+2}^{t-1}\right)\right]$$

$$p_n^{t+1} - p_n^t = \beta\frac{1}{4}\left[p_{n-2}^{t-1} - 2p_{n-1}^{t-1} + 2p_n^{t-1} - 2p_{n+1}^{t-1} + p_{n+2}^{t-1}\right] +$$

$$+(1-\beta)\beta\frac{1}{4}\left[-p_{n-2}^{t-1} + 4p_{n-1}^{t-1} - 6p_n^{t-1} + 4p_{n+1}^{t-1} - p_{n+2}^{t-1}\right] \quad \text{(A2)}$$

The term $p_n^{t+1} - p_n^t$ on the left-hand side is the first order difference with respect to the variable t. Assign this difference with the notation:

$$\Delta p\big|_n^{t+\Delta t} = p_n^{t+1} - p_n^t$$

There is one term on the right-hand side of Eq. (A2) of the form $p_{n-1}^t - 2p_n^t + p_{n+1}^t$ and another one of the form $p_{n-2}^t - 4p_{n-1}^t + 6p_n^t - 4p_{n+1}^t + p_{n+2}^t$. These expressions are the second and fourth order differences respectively with respect to the space variable, centered at the cell n. Assign those differences with the notation:

$$\Delta^2 p\big|_n^t = p_{n-1}^t - 2p_n^t + p_{n+1}^t$$

and

$$\Delta^4 p\big|_n^t = p_{n-2}^t - 4p_{n-1}^t + 6p_n^t - 4p_{n+1}^t + p_{n+2}^t$$

The Eq. (A2) can now be written in terms of the first and second order differentials:

$$\Delta p_n^{t+\Delta t} = \beta\left\{\frac{1}{4}\left[\Delta^2 p_{n-1}^{t-\Delta t} + \Delta^2 p_{n+1}^{t-\Delta t}\right] - (1-\beta)\frac{1}{4}\Delta^4 p_n^{t-\Delta t}\right\} \quad \text{(A3)}$$

For $p(x)$ sufficiently differentiable, the first order differences referred to $x_n$ and $x_{n+1}$ respectively differ by a quantity of order $\Delta x^2$ where $\Delta x = |x_{n+1} - x_n|$. Certainly for $p(x,t)$ the previous calculation is true provided that time is kept constant, that is:

$$\Delta p_{n+1}^t = \Delta p_n^t + O(\Delta x)$$

In a similar way it can be shown that the deviation of the second order differences centered at points $x_{n-1}$ and $x_n$ respectively is of the order of $\Delta x^3$, that is $O(\Delta x^2)$. That is:

$$\Delta^2 p_{n+1}^t = \Delta^2 p_n^t + O(\Delta x^2)$$

Introducing these expressions in (A3 ) we obtain after some simple operations:

$$\frac{\Delta p_n^{t+\Delta t}}{\Delta t}\Delta t = \beta\left\{\Delta x^2\left[\frac{1}{2}\frac{\Delta^2 p_n}{\Delta x^2} + \frac{O[\Delta x^2]}{\Delta x^2}\right] - \Delta x^4(1-\beta)\frac{1}{4}\frac{\Delta^4 p_n}{\Delta x^4}\right\}^{t-\Delta t} \quad \text{(A4)}$$

Or:

$$\frac{\Delta p_n^{t+\Delta t}}{\Delta t} = \beta\left\{\frac{\Delta x^2}{\Delta t}\left[\frac{1}{2}\frac{\Delta^2 p_n}{\Delta x^2} + \frac{O[\Delta x^2]}{\Delta x^2}\right] - \frac{\Delta x^4}{\Delta t}(1-\beta)\frac{1}{4}\frac{\Delta^4 p_n}{\Delta x^4}\right\}^{t-\Delta t} \quad \text{(A5)}$$

Where $\Delta^2 p_n$ and $\Delta^4 p_n$ are the second and fourth order difference expressions. Define:

$$\frac{1}{2}\frac{\Delta x^2}{\Delta t} = D$$

$$\frac{1}{4}\frac{\Delta x^4}{\Delta t} = R$$

Substituting the above relations in (A5):

$$\frac{\Delta p_n^{t+\Delta t}}{\Delta t} = \beta\left\{D\frac{\Delta^2 p_n}{\Delta x^2} + \frac{O\left(\Delta x^2\right)}{\Delta x^2} - \left(1-\beta\right)R\frac{\Delta^4 p_n}{\Delta x^4}\right\}^{t-\Delta t} \qquad \text{(A6)}$$

Taking the limits $\Delta x \to 0$ and $\Delta t \to 0$ and assuming $p(x,t)$ sufficiently smooth the fourth order diffusion equation is obtained:

$$\frac{\partial p}{\partial t} = \beta\left\{D\frac{\partial^2 p}{\partial x^2} - \left(1-\beta\right)R\frac{\partial^4 p}{\partial x^4}\right\}$$